\documentclass[aps,prd,nofootinbib,onecolumn,a4paper,superscriptaddress]{revtex4-2}
\usepackage{bm,latexsym,amsmath,amssymb,amsfonts,mathrsfs,amsfonts}
\usepackage[dvipdfmx]{graphicx}
\usepackage[hang,small,bf]{caption}
\usepackage[subrefformat=parens]{subcaption}
\usepackage{ascmac}
\usepackage{physics}
\usepackage{amsmath}
\usepackage{color,float}
\usepackage{braket}
\usepackage{bbm}
\usepackage{comment}
\usepackage{here}
\usepackage{indentfirst}
\newcommand{\simgt}{\lower.5ex\hbox{$\; \buildrel > \over \sim \;$}}
\newcommand{\simlt}{\lower.5ex\hbox{$\; \buildrel < \over \sim \;$}}

\begin{document}
\title{\bf 
Theoretical Study of the 
Squeezed-Light-Enhanced Sensitivity to Gravity-Induced Entanglement via Finite-Time Analysis
}
\author{Kosei Hatakeyama}
\email{hatakeyama.kosei.103@s.kyushu-u.ac.jp}
\affiliation{Department of Physics,  Kyushu University, 744 Motooka, Nishi-Ku, Fukuoka 819-0395, Japan}
\author{Daisuke Miki}
\affiliation{
The Division of Physics, Mathematics, and Astronomy, California Institute of Technology, Pasadena, CA 91125, USA}
\author{Kazuhiro Yamamoto}

\affiliation{Department of Physics,  Kyushu University, 744 Motooka, Nishi-Ku, Fukuoka 819-0395, Japan}
\affiliation{\small\it Quantum and Spacetime Research Institute, Kyushu University, 744 Motooka, Nishi-Ku, Fukuoka 819-0395 Japan}
\begin{abstract}
We investigate the advantage of using squeezed input light for generating gravity-induced entanglement (GIE) through Fourier-domain analysis.
Based on the findings of Ref.~\cite{Miki2024}, which demonstrated the feasibility of detecting GIE in optomechanical systems under quantum control, we further demonstrate that squeezed input light can reduce the optical noise in the mechanical conditional state and enhance GIE. Furthermore, we estimate the systematic and statistical errors in the measurement of GIE using the Fourier transformation over a finite measurement time. 
Based on the error estimations using the signal-to-noise ratio (SNR) in GIE detection, we find that a total measurement time of $10^6\,\mathrm{s}$ is required to achieve ${\rm SNR} = 1$ when using squeezed input light, whereas $10^{6.8}\,\mathrm{s}$ is needed without squeezed input light. This result highlights the effectiveness of optomechanical systems and the critical role of squeezed input light in enhancing the detectability of GIE.
\end{abstract}
\maketitle
\section{Introduction}
The theory of general relativity has been validated by numerous observations and experiments. 
Notable examples include the direct detection of gravitational waves and observations of black holes, confirming the theory's consistency with empirical data. Despite these successes,
quantum gravity---a unified theory that reconciles gravity with quantum mechanics---remains one of the most significant unresolved problems in modern physics. A major challenge in developing quantum gravity lies in the absence of experimental verification of phenomena arising from the quantum nature of the gravitational field. Such verification requires a regime in which gravitational theory, quantum mechanics, and relativity simultaneously play essential roles. It has long been believed that probing this regime demands extremely high energies, far beyond the capabilities of current experimental technology.
However, recent advancements have opened a promising new possibility for testing the quantum nature of gravity at low energy scales. In particular, two independent groups
---one led by Bose et al., and the other by Marletto and Vedral---have independently proposed an experiment (commonly referred to as the quantum gravity-induced entanglement of masses proposal) to test the quantum superposition of the Newtonian gravitational potential \cite{Bose2017, Marletto2017}. 
These proposals are viewed as a modern revival of Feynman’s famous thought experiment, in which he considered the consequences of a superposition of the gravitational field or space-time  \cite{Feynman1995}. The central concept of these proposals is gravity-induced entanglement (GIE).
According to quantum information theory, entanglement is a nonlocal correlation that cannot be generated by classical processes \cite{Horodecki2009}.
Hence, the observation of GIE would serve as evidence that gravity obeys the principles of quantum mechanics.\footnote{
There are arguments about the consequence of the verification of entanglement by Newtonian gravity, i.e.,  what the GIE means for the quantization of the dynamical gravitational field. 
For example, J. W. Hall and M. Reginatto claim that the BMV experiment does not necessarily mean the existence of a quantized gravitational mediating field suggested by Koopmantype dynamics \cite{Hall2018}.
A similar argument was claimed by C. Anastopoulos and B.-L. Hu \cite{Anastopoulos2022}. 
We must be careful for what the verification of GIE by Newtonian gravity means for quantum gravity, however, it is true that GIE is a natural prediction of the linearized quantum field theory of quantum gravity.
Carney discussed that the gravity-induced entanglement can be a verification of quantized dynamical degrees of freedom in gravitational field theory
under the assumptions of unitarity and Lorentz invariance \cite{Carney22}.
}

Many authors have investigated the method for verifying GIE 
(see e.g., \cite{Miki2024,Bose2017,Marletto2017,Krisnanda2020,Bose2020, Matsumura-Yamamoto2020, Kaku2023, Miki2021, Miki2022, PanLi2023, Zhang2024, Li2024, CarneyPRX2022,Tang2025} for the references).
The experimental verification of GIE remains an open challenge, as no observation has yet been reported.
Due to the weakness of the gravitational interaction, the use of more massive objects is advantageous for measuring gravitational force and GIE. 
Optomechanics, which consists of a movable mirror coupled with an optical mode, is a promising system for generating quantum states of massive objects using feedback control and quantum optimal filtering.
Across many areas of physics, engineering, quantum optics, materials science, and other interdisciplinary fields, various optomechanical systems are rapidly advancing.
The LIGO groups have reported generating a near-quantum state in mirrors with a mass on the order of 10 kilograms \cite{Whittle:2021mtt}. 
Furthermore, the quantum state of a milligram-scale mirror is expected to be realized in the near future \cite{Michimura2020, Matsumoto-HighQ, Matsumoto2019}.
These advancements open opportunities for testing the quantum nature of gravity in optomechanical systems.
The authors of Refs.~\cite{Miao2020,Miao2021,Liu2023,Liu2025,Gerow2024,Miki2025} discussed the detection of the optical signatures induced by gravity. 
Refs.~\cite{Miki2024,Miki2024-2} derived conditions for generating GIE between two mechanical mirrors in optomechanical systems under quantum control, which require that the gravitational effects dominate over both thermal and radiation pressure noises. 
One difficulty in detecting GIE in optomechanical systems lies in distinguishing the signals produced by quantum gravity from those produced by semiclassical gravity, as pointed out in Refs. \cite{Liu2023,Liu2025}. 
They showed that under continuous measurements, the optical squeezing and cross spectra induced by semiclassical and quantum gravity are almost identical. To distinguish the signals, protocols using pulsed optomechanics \cite{Gerow2024} and time-delayed and non-stationary measurements \cite{Miki2025} are proposed. 
Using these protocols, it has been shown that the signals are distinguishable if low-frequency, high-quality-factor mechanical oscillators are prepared.
Another challenge in detecting GIE is to suppress  noises so that the quantum signature of gravity is observable.

It has been shown that squeezed input light reduces radiation pressure noise~\cite{Caves1981, Kimble2001, Helge2009, LIGO-squeezing, Grangier1987, Xiao1987}. 
Extending the previous study~\cite{Miki2024}, we demonstrate in the present paper that the use of squeezed input light enhances GIE and relaxes the experimental requirements for its generation.
Furthermore, we investigate the systematic and statistical errors due to the finite measurement time. 
In a realistic experiment, the finite duration of measurements introduces systematic and statistical errors that were absent in the idealized case of infinite measurement time considered in Ref.~\cite{Miki2024}. 
These errors can affect the amount of GIE and the time required for detection since the effects of gravity are significantly small.
We evaluate the signal-to-noise ratio (SNR) under the finite measurement time and show the reduction in the optical input noise due to  squeezed input enhances the SNR of GIE and shortens the time required to achieve SNR $= 1$.

This paper is organized as follows: 
In Sec.~II, we briefly review the optomechanical systems from our previous work \cite{Miki2024} and introduce the causal quantum Wiener filter, including the formulas for squeezed input light.
Sec.~III examines GIE between two mirrors in the Fourier domain and determines the optimal squeezing parameters for {\color{red}GIE} generation.
In Sec. IV, we analyze the systematic and statistical errors associated with finite measurement time. We also evaluate the total measurement time required to achieve ${\rm SNR} = 1$, taking these errors into account.
Sec.~V summarizes our findings and presents the conclusions.
Appendix A provides the mathematical expressions for the degree of entanglement in the Fourier domain using the quantum Wiener filter and squeezed input light.
Appendix B explains why the characteristic error‑suppression time scales as $1/\gamma_m$, where $\gamma_m$ is the mirror’s effective mechanical damping rate.
In Appendix C, the feedback gain is estimated by considering the feedback noise, ensuring that it remains negligible relative to the thermal noise.
\section{FORMULAS}
\begin{figure}[t]
\begin{center}
\includegraphics[width=130mm]{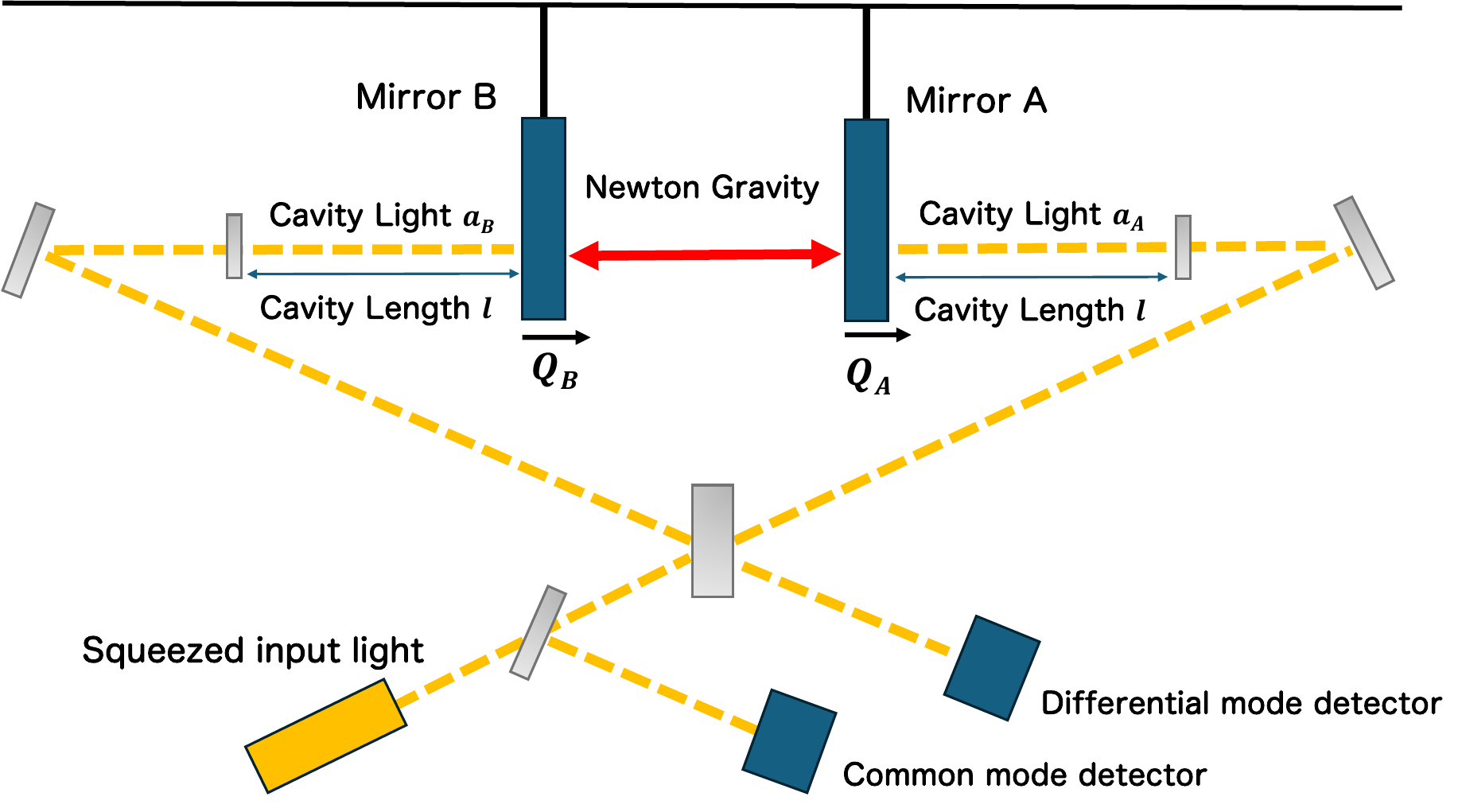}
\caption{Configuration of the optomechanical system of the symmetrical setup. 
See also the footnote in the next page for "squeezed input light". 
\label{Setup}}
\end{center}
\end{figure}

We consider a cavity-optomechanical system consisting of two mechanical mirrors, each of mass $m$, as illustrated in Fig. \ref{Setup}. The total Hamiltonian is
\begin{align}
    \hat{H}&=\frac{\hat{P}^2_A}{2m}+\frac{1}{2}m\Omega^2\hat{Q}^2_A+\frac{\hat{P}^2_B}{2m}+\frac{1}{2}m\Omega^2\hat{Q}^2_B+\hbar\omega_c\hat{a}^{\dagger}_A\hat{a}_A+\frac{\hbar\omega_c}{\ell}\hat{Q}_A\hat{a}^{\dagger}_A\hat{a}_A+\hbar\omega_c\hat{a}^{\dagger}_B\hat{a}_B-\frac{\hbar\omega_c}{\ell}\hat{Q}_B\hat{a}^{\dagger}_B\hat{a}_B\nonumber\\
    &+i\hbar E\Big(e^{-i\omega_Lt}\frac{\hat{a}^{\dagger}_A+\hat{a}^{\dagger}_B}{\sqrt{2}}-e^{i\omega_Lt}\frac{\hat{a}_A+\hat{a}_B}{\sqrt{2}}\Big)-\frac{Gm^2}{L}\frac{1}{1-(\hat{Q}_B-\hat{Q}_A)/L},
    \label{Hamiltonian}
\end{align}
where $\hat{Q}_j$ and $\hat{P}_j$ with $j=A, B$ are the position and momentum operators of the two mirrors satisfying $[\hat{Q}_j, \hat{P}_j]=i\hbar$.
$\hat{a}_j$ and $\hat{a}^{\dagger}_j$ with $j=A, B$ are the annihilation and creation operators of optical cavity modes, with $[\hat{a}_j,\hat{a}^{\dagger}_j]=1$. 
$\omega_c$ and $\ell$ are the cavity resonance frequency and length, respectively. 
$E=\sqrt{P_{\rm in}\kappa/\hbar\omega_L}$ is the input laser amplitude, where $P_{in}$ is the laser power, $\kappa$ is the cavity decay rate.
The terms on the right-hand side of the first line of Eq.~(\ref{Hamiltonian}) represent the Hamiltonian for the mirrors and the cavity field, as well as the coupling between them via radiation pressure \cite{Aspelmeyer2014}.

The first term on the second line corresponds to the Hamiltonian for the cavity driven by an input laser at frequency $\omega_L$, modeled using the Gardiner-Collett framework~\cite{Gardiner-Collett1985}, where we consider the resonant case $\omega_L\simeq\omega_c$. 
The final term describes the gravitational interaction between the two mirrors.

Under linearization and the neglect of higher-order contributions in the subsequent perturbative expansion, the interaction between mirrors A and B uniquely reduces to $-2Gm^2/L^3\hat{Q}_A\hat{Q}_B$. 
It is assumed that the coupling between the two mirrors is exclusively gravitational.

The Coulomb interaction produces the same type of bilinear coupling $\hat Q_A\hat Q_B$ as gravity; therefore, it is essential to make the mirrors as electrically neutral as possible. The magnitudes of the gravitational and Coulomb potential energies are given by $|\phi_{\rm Grav.}|=Gm^2/L$ and $|\phi_{\rm Coul.}|=N^2e^2/(4\pi\varepsilon_0 L)$, respectively, where $N$ denotes the number of excess elementary charges. Their ratio,
$|\phi_{\rm Grav.}/\phi_{\rm Coul.}|=3\times10^{11}/N^2~(m/1~{\rm g})^2$, shows that the Coulomb interaction becomes negligible as long as $N\simlt6\times10^5$ for mirrors of mass $m=1~\text{g}$.
Even for electrically neutral mirrors, electromagnetic interactions such as the Casimir–Polder force can degrade the entanglement between the two objects \cite{Kamp2020,Schut2024,Fragolino2024}.

Ref.~\cite{Miki2024} shows that the Casimir–Polder contribution can be neglected for mirrors with $m\gg10^{-7}~\text{kg}$ and separations comparable to their size. In addition, it has been pointed out that apparent correlations may arise even within a semiclassical description of gravity when the mirrors are continuously monitored \cite{Liu2023,Liu2025}. However, such correlations can be distinguished from those generated by quantum gravity through time-delayed or non-stationary measurement schemes \cite{Miki2025,Liu2025-2}.

We consider the perturbations around a steady state by writing $\hat{Q}_j=\Bar{Q}_j+\delta\hat{Q}_j$, $\hat{P}_j=\Bar{P}_j+\delta\hat{P}_j$, $\hat{a}_j=\Bar{a}_j+\delta\hat{a}_j$, $\hat{a}^{\dagger}_j=\Bar{a}^{*}_j+\delta\hat{a}^{\dagger}_j$, where $(\Bar{Q}_j, \Bar{P}_j ,\Bar{a}_j)$ are the classical mean values of $(\hat{Q}_j, \hat{P}_j ,\hat{a}_j)$ and $(\delta\hat{Q}_j, \delta\hat{P}_j ,\delta\hat{a}_j)$ are the perturbative quantities. 
We then expand the Hamiltonian to the second order of the perturbations. 
We introduce the dimensionless common (+) and differential (-) mode as
\begin{align}
\delta\hat{q}_{\pm}=\sqrt{\frac{2m\Omega_{\pm}}{\hbar}}\frac{\delta\hat{Q}_A\pm\delta\hat{Q}_B}{\sqrt{2}},~~~
\delta\hat{p}_{\pm}=\sqrt{\frac{2}{m\hbar\Omega_{\pm}}}\frac{\delta\hat{P}_A\pm\delta\hat{P}_B}{\sqrt{2}},~~~
\delta\hat{a}_{\pm}=\frac{\delta\hat{a}_A\mp\delta\hat{a}_B}{\sqrt{2}},
\end{align}
which satisfy $[\hat{q}_{\pm}, \hat{p}_{\pm}] = 2i$.
Here  we defined $\Omega_+=\Omega,~~\Omega_-=\Omega\sqrt{1-\epsilon}$, where $\epsilon={4Gm}/({L^3\Omega^2})=4G\rho\Lambda/\Omega^2$ describes the gravitational coupling. $\rho$ and $\Lambda$ are the mass density and the form factor of the mirror, respectively \cite{Miao2020}.
In the present paper, the mirror material is assumed to be a platinum-gold alloy ($\rho=20 [\text{g/cm}^3]$), and the shape factor is assumed to be $\Lambda=2$, following Refs.~\cite{Miao2020,Miki2024}. Under these parameters, the coupling is $\epsilon=0.27$ for $\Omega/2\pi=10^{-3}~{\rm Hz}$.

The Langevin equations for the perturbations are written as
\begin{align}
    \dot{\delta\hat{q}}_{\pm}&=\Omega_{\pm}\delta\hat{p}_{\pm},\\
    \dot{\delta\hat{p}}_{\pm}&=-\Omega_{\pm}\delta\hat{q}_--2g_{\pm}\delta\hat{x}_{\pm}-\gamma_m\delta\hat{p}_{\pm}+\sqrt{2\gamma_m}\hat{p}^\mathrm{in}_{\pm},\\
    \dot{\delta\hat{x}}_{\pm}&=-\frac{\kappa}{2}\delta\hat{x}_{\pm}+\sqrt{\kappa}\hat{x}^\mathrm{in}_{\pm},\\
    \dot{\delta\hat{y}}_{\pm}&=-2g_{\pm}\delta\hat{q}_{\pm}-\frac{\kappa}{2}\delta\hat{y}_{\pm}+\sqrt{\kappa}\hat{y}^\mathrm{in}_{\pm},
\end{align}
where $\gamma_m$ is the effective mechanical decay rate under feedback control \cite{genes}, and $\kappa$ is the photon decay rate in the cavity field. 
The terms on the right-hand side, $\hat p^\mathrm{in}_{\pm}, \hat x^\mathrm{in}_{\pm}$, and $\hat y^\mathrm{in}_{\pm},$ along with those associated with the decay rates, are introduced to model dissipation and noise effects in accordance with the fluctuation-dissipation relation. 
$\hat{p}^\mathrm{in}_{\pm}$ is the thermal fluctuation, $\hat x_\pm^\mathrm{in}$ and $\hat y_\pm^\mathrm{in}$ are the optical input noise.
We note here that the effect of gravity enters only through the frequency of the differential mode $\Omega_-$, which leads to the asymmetry between common/differential modes and thus to the generation of GIE between the mirrors. 
The parameter 
$g_{\pm}$ is the optomechanical coupling, given by
\begin{align}
g_{\pm}=\frac{\omega_c}{\ell}\sqrt{\frac{\hbar}{m\Omega_{\pm}}}\frac{E}{\kappa}.
\end{align}

Defining the variables in the Fourier domain as $\displaystyle{f(\omega)=\int^{\infty}_{-\infty}dtf(t)e^{i\omega t}}$, we obtain the solution of the Langevin equations in the steady state as
\begin{align}
    \delta\hat{q}_{\pm}(\omega)&=\frac{\Omega_{\pm}}{F_{\pm}(\omega)}\Big(\sqrt{2\gamma_m}\hat{p}^\mathrm{in}_{\pm}(\omega)-\frac{4g_{\pm}}{\sqrt{\kappa}}\hat{x}^\mathrm{in}_{\pm}(\omega)\Big),\\
    \delta\hat{p}_{\pm}(\omega)&=-\frac{i\omega}{\Omega_{\pm}}\delta\hat{q}_{\pm}(\omega),\\  
    \delta\hat{x}_{\pm}(\omega)&=\frac{2}{\sqrt{\kappa}}\hat{x}^\mathrm{in}_{\pm}(\omega),\\
    \delta\hat{y}_{\pm}(\omega)&=-\frac{4g_{\pm}}{\kappa}\delta\hat{q}_{\pm}(\omega)+\frac{2}{\sqrt{\kappa}}\hat{y}^\mathrm{in}_{\pm}(\omega),
\end{align}
where we defined $F_{\pm}(\omega)=\Omega_{\pm}^2-i\gamma_m\omega-\omega^2$, and assumed the adiabatic approximation $\omega \ll \kappa$.
We assume that the correlation functions of the thermal noise are given by 
\begin{align}
   \langle\{\hat{p}^\mathrm{in}_{\pm}(\omega),\hat{p}^\mathrm{in}_{\pm}(\omega')\}\rangle&=2(2n^{\pm}_\mathrm{th}+1)2\pi\delta(\omega+\omega'),
\end{align}
where $\{\hat{A},\hat{B}\}=\hat{A}\hat{B}+\hat{B}\hat{A}$ is the anti-commutator, and $n^{\pm}_\mathrm{th}=k_BT_0\Gamma/(\hbar\Omega_{\pm}\gamma_m)$ is the thermal phonon number under feedback, $k_B$ is Boltzmann's constant, and $T_0$ is the environmental temperature.
The previous work demonstrated that GIE can be generated between mirrors in optomechanical systems under quantum control and identified the conditions for its generation in terms of the Fourier modes of the mechanical motion \cite{Miki2024}. 
In this section, we extend the formulation by incorporating squeezed input light, which serves to reduce noise and enhance the signal.\footnote{
We assume that the quantum perturbative components of squeezed input light are those of a squeezed vacuum state. This corresponds to injecting squeezed vacuum light through the output port  of the interferometer, as proposed in Ref.~\cite{Kimble2001}. Since our setup is designed to detect gravity-induced entanglement (GIE), special care must be taken in practical implementations to avoid introducing asymmetries in the squeezing parameters between the common and differential modes.
}
The extension to the case of squeezed input light is obtained by assuming
\begin{align}
    \langle\{\hat{x}^\mathrm{in}_{\pm}(\omega),\hat{x}^\mathrm{in}_{\pm}(\omega')\}\rangle&=2(2N_\mathrm{th}+1)(\cosh{2r}+\cos{2\phi}\sinh{2r})\times2\pi\delta(\omega+\omega'),\\
    \langle\{\hat{y}^\mathrm{in}_{\pm}(\omega),\hat{y}^\mathrm{in}_{\pm}(\omega')\}\rangle&=2(2N_\mathrm{th}+1)(\cosh{2r}-\cos{2\phi}\sinh{2r})\times2\pi\delta(\omega+\omega'),\\
    \langle\{\hat{x}^\mathrm{in}_\mathrm{\pm}(\omega),\hat{y}^\mathrm{in}_{\pm}(\omega')\}\rangle&=2(2N_\mathrm{th}+1)\sin{2\phi}\sinh{2r}\times
 2\pi\delta(\omega+\omega'),
    \label{xinyin}
\end{align}
to reproduce the correlation function of the optical fluctuation in the squeezed vacuum state \cite{Helge2009},
where $r$ is the optical squeezing strength, $\phi\in[0,\pi]$ is the squeezing angle, and $N_\mathrm{th}=(e^{\hbar\omega_c/k_BT_0}-1)^{-1}$ is the thermal photon number, which can be regarded as $N_\mathrm{th}=0$ in the present paper because of the high optical frequency $\hbar\omega_c/k_BT_0\gg1$.
The squeezed angle $\phi$ is assumed to be independent of the mechanical frequency $\Omega$. Unlike the vacuum input {\color{blue}($r=0$)}, the correlation of the $\hat x_\pm^\mathrm{in}$ and $\hat y_\pm^\mathrm{in}$ in Eq.~\eqref{xinyin} can be non-zero for the squeezed input. 
If this correlation is negative, we can reduce the optical noise.
In the next section, it is shown that GIE can be enhanced by appropriately choosing $\phi$.

The input-output relation of the phase quadrature is given by (see e.g., \cite{Aspelmeyer2014})
\begin{align}
 \hat{Y}_{\pm}(\omega)&=\hat{y}_{\pm}^\mathrm{in}(\omega)-\sqrt{\kappa}\delta\hat{y}_{\pm}(\omega)
 =\frac{4g_{\pm}}{\sqrt{\kappa}}\hat{q}_{\pm}(\omega)-\hat{y}^\mathrm{in}_{\pm}(\omega). 
\end{align}
In this analysis, it is assumed that no additional vacuum noise is introduced into the optical output quadrature due to measurement imperfections.
We introduce the spectrum density $S^\pm_{AB}(\omega)$, which is defined as follows;
\begin{align}
    2\pi\delta(\omega-\omega')S^\pm_{AB}(\omega)=\frac{1}{2}\langle\{\hat{A}_\pm(\omega), \hat{B}_\pm^{\dagger}(\omega')\}\rangle,
\end{align}
where $A$ denotes $Y$, $q$, and $p$, and $B$ denotes $Y$.
Explicitly, the mechanical and optical spectral densities are given by
\begin{align}
    S^{\pm}_{YY}(\omega)&=\frac{32\gamma_mg_{\pm}^2\Omega_{\pm}^2}{\kappa|F_{\pm}(\omega)|^2}(2n_\mathrm{th}^{\pm}+1)+\frac{256g_{\pm}^4\Omega_{\pm}^2}{\kappa^2|F_{\pm}(\omega)|^2}(2N_\mathrm{th}+1)(\cosh{2r}+\cos{2\phi}\sinh{2r})\nonumber\\
    &~~+(2N_\mathrm{th}+1)(\cosh{2r}-\cos{2\phi}\sinh{2r})+\frac{16g_{\pm}^2\Omega_{\pm}}{\kappa}\Big(\frac{1}{F_{\pm}(\omega)}+\frac{1}{F^*_{\pm}(\omega)}\Big)(2N_\mathrm{th}+1)\sin{2\phi}\sinh{2r}.
    \\
    S^{\pm}_{qY}(\omega)&=\frac{8g_{\pm}\Omega_{\pm}^2\gamma_m}{\sqrt{\kappa}|F_{\pm}(\omega)|^2}(2n_\mathrm{th}^{\pm}+1)+\frac{64g_{\pm}^3\Omega_{\pm}^2}{\kappa\sqrt{\kappa}|F_{\pm}(\omega)|^2}(2N_\mathrm{th}+1)(\cosh{2r}+\cos{2\phi}\sinh{2r})\nonumber\\
    &~~+\frac{4g_{\pm}\Omega_{\pm}}{\sqrt{\kappa}F_{\pm}(\omega)}(2N_\mathrm{th}+1)\sin{2\phi}\sinh{2r},\\
    S^{\pm}_{pY}(\omega)&=-\frac{i\omega}{\Omega_{\pm}}S^{\pm}_{qY}.
\end{align}
Next, we consider the quantum Wiener filter \cite{Miki2024,Shichijo2023, JMY}, which estimates the mechanical motion conditioned on the measurement results, and we define the conditioned mechanical quadratures as
\begin{align}
   & \tilde{q}_{\pm}=\delta\hat{q}_{\pm}(\omega)-H^{\pm}_{q}(\omega)\hat{Y}_{\pm},\qquad \tilde{p}_{\pm}=\delta\hat{p}_{\pm}(\omega)-H^{\pm}_{p}(\omega)\hat{Y}_{\pm},
\end{align}
where the causal Wiener filter functions $H_{q}^\pm$ and $H_{p}^\pm$ are defined by
\begin{align}
    \label{Hq}
    &H^{\pm}_{q}:=\frac{1}{S^{\pm\mathrm{C}}_{YY}(\omega)}\Bigg\lbrack \frac{S^{\pm}_{qY}(\omega)}{S^{\pm\mathrm{NC}}_{YY}(\omega)}\Bigg\rbrack_+=\frac{\sqrt{\kappa}}{4g_{\pm}}\frac{\omega_{Y}^{\pm2}-\Omega_{\pm}^2-i\omega(\gamma_{Y}^\pm-\gamma_m)}{F'_\pm(\omega)},~~~~~~~\\
    &H^{\pm}_{p}:=\frac{1}{S^{\pm\mathrm{C}}_{YY}(\omega)}\Bigg\lbrack \frac{S^{\pm}_{pY}(\omega)}{S^{\pm\mathrm{NC}}_{YY}(\omega)}\Bigg\rbrack_+=\frac{\sqrt{\kappa}}{4g_{\pm}\Omega_{\pm}}\frac{-(\gamma_{Y}^\pm-\gamma_{m})\Omega_{\pm}^2-i\omega(\omega_{Y}^{\pm2}-\Omega_{\pm}^2-(\gamma_{Y}^\pm-\gamma_m)\gamma_m)}{F'_\pm(\omega)},
    \label{Hp}
\end{align}
where $F'_\pm(\omega):=\omega_{Y\pm}^2-i\gamma_{Y\pm}\omega-\omega^2$.
Here, the subscript $\mathrm{C}$ and $\mathrm{NC}$ mean the causal and non-causal part, and $[...]_+$ represents the causal function of the spectrum density. The causal part can be calculated via Wiener–Hopf factorization $S_{ZY}=S_{ZY}^{\mathrm {C}}\cdot S_{ZY}^{\mathrm{NC}}$, where $Z$ denotes $q$, $p$, and $Y$.
The causal function is a partial fractional decomposition of the denominator of the spectrum density into causal/non-causal parts, and the causal part is selected.

Following the Ref.~\cite{JMY}, the extended forms of $\gamma_{Y}^\pm$, $\omega_{Y}^\pm$, $\lambda_{\pm}$ and $\Lambda_{\pm}$ are given by,
\begin{align}
    \gamma_{Y}^\pm&=\sqrt{\gamma_m^2-2\Omega_{\pm}(\Omega_{\pm}+\Lambda_{\pm})+2\omega_{Y}^{\pm2}}~,\nonumber\\
    \omega_{Y}^\pm&=\sqrt{\Omega_{\pm}\sqrt{\Omega_{\pm}^2+2\Lambda_{\pm}\Omega_{\pm}+\Big(2\gamma_m(2n_\mathrm{th}^{\pm}+1)+\frac{16g_{\pm}^2}{\kappa}(2N_\mathrm{th}+1)(\cosh{2r}+\cos{2\phi}\sinh{2r}\Big)\lambda_{\pm}}}~,\nonumber\\
    \lambda_{\pm}&=\frac{16g_{\pm}^2}{\kappa(2N_\mathrm{th}+1)(\cosh{2r}-\cos{2\phi}\sinh{2r})}~,\quad\Lambda_{\pm}=\frac{16g_{\pm}^2}{\kappa}\frac{\sin{2\phi}\sinh{2r}}{(\cosh{2r}-\cos{2\phi}\sinh{2r})}.\nonumber
\end{align}
The quantities $\gamma^\pm_Y$ and $\omega^\pm_Y$ correspond to the damping rate and resonance frequency apparently caused by Wiener filtering, respectively.
$\lambda_\pm$ represents the measurement rate, i.e., the inverse time  required to resolve the zero-point motion.
An increase in $\lambda_\pm$ reduces the conditional position variance, meaning that the $q$-squeezed state is more enhanced.
Furthermore, due to the use of squeezed input light, $\Lambda_\pm$ remains non-zero even at $\Delta=0$, in contrast to the result in Ref.~\cite{JMY}.
\section{Effects of squeezed input light on GIE}
Next, we focus on the GIE between the two mirrors generated in the steady state. Following the previous work \cite{Miki2024}, we introduce the degree of entanglement in the Fourier domain, with the entanglement criterion is given by
\begin{align}
    \langle E_{\rm Fil}(\omega)\rangle := \frac{\langle\hat{R}_{\tilde{q}_+}(\omega)^2\rangle\langle\hat{R}_{\tilde{p}_-}(\omega)^2\rangle\Omega_-/\Omega}{|\langle\lbrack\hat{R}_{\tilde{q}_A}(\omega),\hat{R}_{\tilde{p}_A}(\omega)\rbrack\rangle|^2}<1,
    \label{E(w)}
\end{align}
where $\hat{R}_Z(\omega)$ is defined by $\hat{R}_Z(\omega)=(\hat{Z}(\omega))+\hat{Z}(-\omega))/2$ for $Z=\tilde{q}_+,~\tilde{p}_-,\tilde{q}_A,~\tilde{p}_A$. 
We note that $\hat{R}_Z(\omega)$ is the Hermitian operator in the Fourier domain, and that  
 $\tilde{q}_A(\omega)=(\tilde{q}_+(\omega)+\tilde{q}_-(\omega)\sqrt{\Omega/\Omega_-})/2$ and $\tilde{p}_A(\omega)=(\tilde{p}_+(\omega)+\tilde{p}_-(\omega)\sqrt{\Omega_-/\Omega})/2$ are the conditional canonical operators of the mechanical mirror.
The above entanglement criterion was originally introduced in Ref.~\cite{Mancini2002}.
The subscript "Fil" in $\langle E_{\rm Fil}(\omega)\rangle$ indicates that the quantities $\tilde{q}_\pm$ and $\tilde{p}_\pm$ are estimated using the quantum Wiener filter
(See \cite{Miki2024} for the details).
Eq.~(\ref{E(w)}) provides a sufficient condition for  entanglement. The denominator of $\langle E_{\rm Fil}(\omega)\rangle$ is determined by the commutation relations of noise operators,
$    \lbrack\hat{p}^\mathrm{in}_{\pm}(\omega),\hat{p}^\mathrm{in}_{\pm}(\omega')\rbrack=({2\omega}/{\Omega_{\pm}})\times2\pi\delta(\omega+\omega'),$ 
and $\lbrack\hat{x}^\mathrm{in}_{\pm}(\omega), \hat{y}^\mathrm{in}_{\pm}(\omega') \rbrack=2i\times2\pi\delta(\omega+\omega')$, where $\lbrack\hat{A}, \hat{B}\rbrack=\hat{A}\hat{B}-\hat{B}\hat{A}$ denotes the commutator.
\begin{table}[t]
    \centering
    \begin{tabular}{|c|c|c|c|c|}\hline
Symbol & &Description & Value & Unit \\ \hline
$\Omega/2\pi$ & &Frequency of the mirror & $10^{-3}$ & $\rm s^{-1}$ \\ \hline
$\omega_c/2\pi$ & &Resonance frequency of the cavity lights & $2.818\times10^{14}$ & $\rm s^{-1}$ \\ \hline
$\omega_L/2\pi$ & &Frequency of the input laser& $2.818\times10^{14}$ & $\rm s^{-1}$ \\ \hline 
$m$ & &Mirror mass & $1$ & g \\ \hline
$\ell$ & &Cavity length & $10^{-1}$ & m \\ \hline
$T_0$ & &Environment Temperature & $1$ & K \\ \hline
$\gamma_m/2\pi$ &  &Mechanical decay rate under feedback control & $6.6\times10^{-6}$ & $\rm s^{-1}$ \\ \hline
$\Gamma/2\pi$ & &Mechanical dissipation rate & $10^{-18}$ & $\rm s^{-1}$ \\ \hline
$\kappa/2\pi$ & &Cavity decay rate & $10^{8}$ & $\rm s^{-1}$ \\ \hline
    \end{tabular}
    \caption{Parameters used in the present study.    }
    \label{table1}
\end{table}

\begin{figure}[t]
   \begin{center}
\includegraphics[width=80mm]{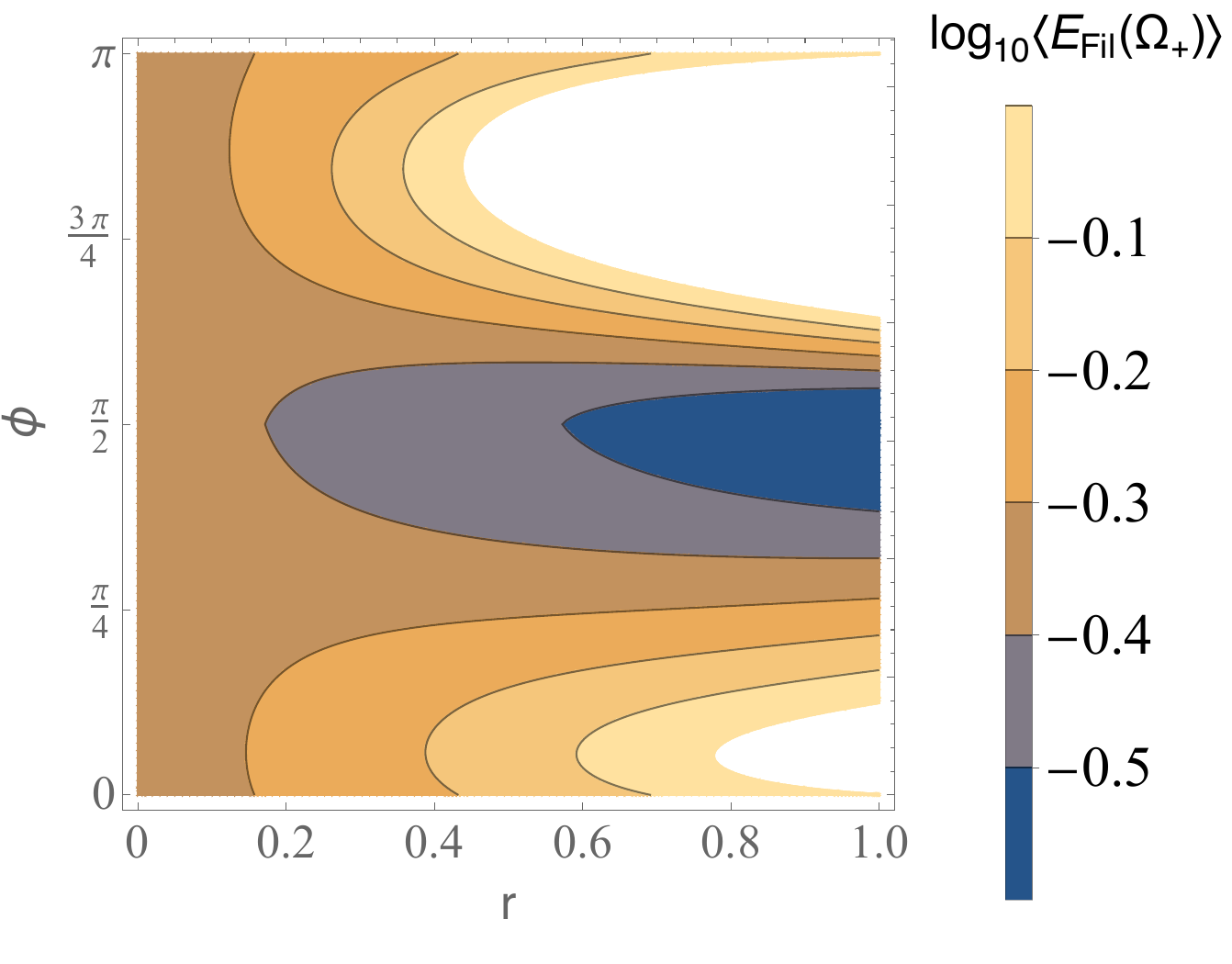}
   \hspace{.30cm}
\includegraphics[width=80mm]{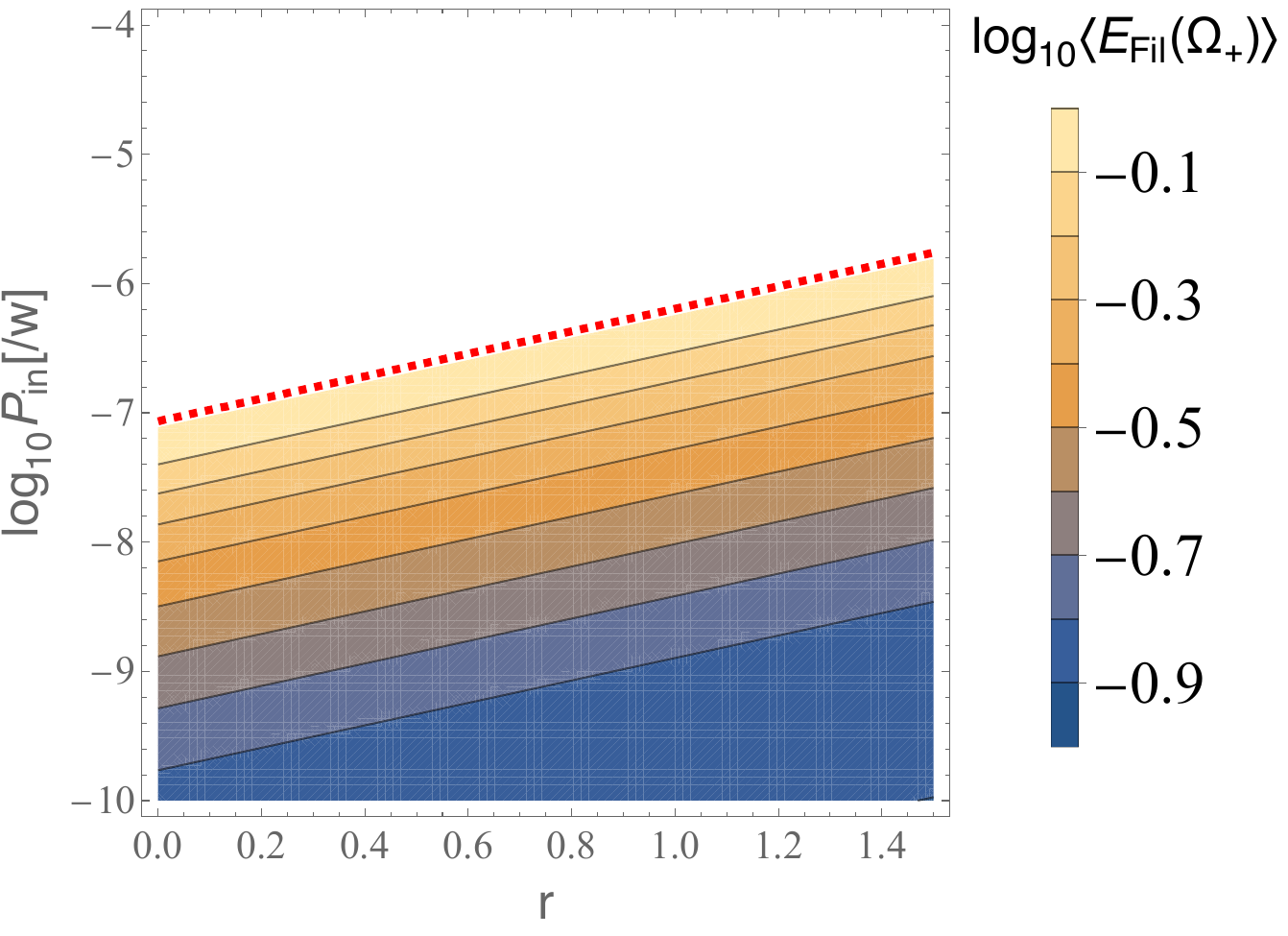}
\caption{The left panel shows the contour plot of \( \log_{10}\langle E_{\rm Fil}(\Omega_+)\rangle \) on the plane of the squeezing parameters \( (r, \phi) \), while the right panel shows the same quantity plotted on the plane of \( (r, \log_{10}P_{\mathrm{in}}~[\mathrm{W}]) \).
In the left panel, the input laser power is fixed at \( P_{\mathrm{in}} = 10^{-10} \) [W], while in the right panel, the squeezing angle is set to \( \phi = \pi/2 \).  
The white regions indicate that the two mechanical mirrors are not entangled.  The red line in the right panel represents the condition given in Eq.~\eqref{condition1}.  
Darker blue regions correspond to a stronger degree of entanglement.  
The left panel shows that increasing \( r \) enhances entanglement when \( \phi \simeq \pi/2 \).  
The right panel demonstrates that increasing \( r \) suppresses radiation-pressure noise, thereby enabling entanglement generation even at higher input laser powers.
}
\label{logE-contour}
  \end{center}
\end{figure}

The left panel of Fig.~\ref{logE-contour} shows the contour of $\log_{10}\langle E_{\text{Fil}}(\Omega_+)\rangle$
on the plane defined by the squeezing parameter $r$ and the squeezing angle $\phi$, with all other parameters fixed as in Table I.
The colored region satisfies the entanglement criterion given in Eq.~(\ref{E(w)}), while the white region does not. 
When $\phi\simeq{\pi /2}$, the entanglement becomes stronger as $r$ increase, as indicated by the decrease in $\langle E_{\text{Fil}}(\Omega_+)\rangle$.
For instance, at $r=1$ and $\phi=\pi/2$, we have $\langle E_{\rm Fil}(\Omega_+)\rangle\simeq0.30$. 
At $\phi=\pi/2$, the phase noise increases by a factor $e^{2r}$, while the amplitude noise decreases by $e^{-2r}$. 
The amplified phase noise, which corresponds to the measurement noise in phase-quadrature detection, can be effectively suppressed using the quantum Wiener filter.
Consequently, the use of squeezed input laser light enhances entanglement when the squeezing angle $ \phi$ is appropriately chosen.
Squeezing the input laser reduces optical noise, improves the purity of the conditional state, and increases the asymmetry between the common and differential modes.
The right panel of Fig.~\ref{logE-contour} shows the contour plot of \( \log_{10}\langle E_{\rm Fil}(\Omega_+)\rangle \) as a function of the laser power \( P_{\mathrm{in}} \) and the squeezing \( r \), by fixing $\phi$ at $\pi/2$.
Increasing the squeezing strength of the input light relaxes the requirement on the optical power.
A strongly entangled state is realized in the darker blue region, where strong squeezing effectively suppresses the optical input noise.

Ref.~\cite{Miki2024} derived simple conditions for generating GIE without the quantum Wiener filter for $\omega=\Omega_+$ and $N_{\rm th}=0$ (see Eqs. (49) and (51) in Ref.~\cite{Miki2024}).
Here, we generalize these conditions to the case with squeezed input light, which can be written in a combined form as
\begin{align}
    Q_{+} \epsilon &> 2(2n_{\mathrm{th}}^++1)+4C_+(\cosh{2r} + \cos{2\phi} \sinh{2r}), \label{condition1} 
\end{align}
where $Q_\pm=\Omega_\pm/\gamma_m$ is the quality factor, $C_\pm=4g_\pm^2/\gamma_m\kappa$ is the cooperativity.
The parameters $Q_\pm$ are determined by the mechanical resonant frequency and the dissipation rate of the system, while $C_{\pm}$ characterizes the strength of the measurement.
The inequality~\eqref{condition1} represents the condition under which the gravitational interaction between the mirrors dominates over the optomechanical coupling and thermal noise.
In particular, it shows that entanglement generation is enhanced by an appropriate choice of the squeezing angle $\phi$.
At $\phi = \pi/2$, entanglement is achievable even at higher input laser powers, as optical squeezing suppresses radiation pressure noise and facilitates entanglement generation.
Since optical squeezing affects only the laser field, it does not mitigate the thermal noise contribution.

In the case with Wiener filtering, the exact form of $E(\omega)$ is complicated (see Appendix~A).  Ref.~\cite{Miki2024} presented only numerical result for the entanglement behavior and did not derive an  explicit entanglement condition with Wiener Filtering.
However, when the squeezed angle $\phi =\pi/2$ is fixed, where the entanglement becomes maximal, we derive the entanglement condition with the Wiener filtering as
\begin{align}
    Q_+\epsilon>2(2n_\mathrm{th}^++1)+4C_+e^{-2r}+\frac{1}{4C_+}\Big(\frac{\gamma_Y^+-\gamma_m}{\gamma_m}\Big)^2e^{2r}-\Big(1+\frac{\omega^{+2}_Y-\Omega^2_+}{\gamma_Y^{+2}}\epsilon\Big)\frac{\gamma_Y^+-\gamma_m}{\gamma_m}.
    \label{condition3}
\end{align}
In deriving the above expression, we used the conditions $(\omega_Y^{+2}-\Omega_+^2)^2\ll\Omega_+^2(\gamma_Y^+-\gamma_m)^2$ and $|H^+_q(\omega_+)|^2\simeq |H^-_p(\omega_+)|^2$, whose validity was checked numerically in the entanglement region. 
The last two terms on the right-hand side represent the filtering effect.  
Around the boundary of the entanglement region, these two terms approximately cancel each other, which explains the slight difference observed between the cases with and without the quantum Wiener filter in the absence of optical squeezing, as shown in Figs.~2 and 3 of Ref.\cite{Miki2024}.
The red dashed line in Fig.~\ref{logE-contour} represents the boundary defined by the inequality~\eqref{condition3}, indicating that the analytical condition provides a good approximation in the parameter regime where the entanglement criterion~\eqref{E(w)} is satisfied.

Furthermore, we should discuss the environmental parameters required to realize GIE. 
Because the gravitational interaction is extremely weak, decoherence effects of all other sources—particularly thermal noise—must be smaller than the gravitational effect. 
Under the conditions \eqref{condition1} and \eqref{condition3}, achieving entanglement at $\epsilon=0.27$ requires an extremely low level of dissipation, quantified by $\Gamma T / 2\pi < 10^{-18}[\mathrm{Hz\cdot K}]$.
Assuming the ideal gas damping with mass $m_{\rm atom}$ and pressure $P$, the dissipation rate for a cylindrical-shaped mirror with radius $R$ and  height $h$ is given by $\Gamma =(PR^2\sqrt{8\pi m_{\rm env}}/m\sqrt{k_BT})(1+(h/2R)+\pi/4)$ \cite{Cavalleri2010,matsumoto2025spacebasedmmmgscalelaserinterferometer}.
For an environment dominated by hydrogen atoms, we have $\Gamma/2\pi\sim10^{-18}~{\rm Hz}~(P/10^{-14}~{\rm Pa})(T/1~{\rm K})^{-1/2}(1~{\rm g}/m)^{1/3}(20~{\rm g}~{\rm cm}^{-3}/\rho)^{2/3}$.
Although the required vacuum level is extraordinarily demanding, laboratory experiments have already achieved $5\times10^{-15}\,\mathrm{Pa}$ at $4.2\,\mathrm{K}$ using cryopumping~\cite{Gabrielse2002}. 
Moreover, BASE experiment at CERN~\cite{Baglin2024,Smorra2015} has estimated a pressure of $10^{-17}\,\mathrm{Pa}$ at $4.5\,\mathrm{K}$, based on measurements of antiproton annihilation events caused by residual gas scattering.

\hspace{1cm}
\section{Error Estimation of degree of entanglement in finite measurement time}
The results of Ref.~\cite{Miki2024} are based on the Fourier transform under the idealized assumption of an infinitely long measurement duration. 
In realistic experimental setups, however, measurements are performed over a finite time interval.
In this section, we extend the analysis by considering the Fourier transform over a finite measurement time $T$. 
Finite measurement time introduces errors that can degrade the precision of GIE estimation, as the GIE signal is inherently weak.
We analyze the impact of finite-time measurements on GIE detection and estimate the minimum measurement duration required to observe the GIE signal.

We introduce the Fourier transform over a finite time duration as
\begin{align}
    \delta\hat{q}_{d\pm}(t)=\sum_{n=-\infty}^{\infty}\delta\hat{q}_{d\pm}(\omega_n)\frac{e^{-i\omega_n t}}{\sqrt{T}},~~~~~~
    \delta\hat{q}_{d\pm}(\omega_n)=\int^{T/2}_{-T/2}dt~\delta\hat{q}_{d\pm}(t)\frac{e^{i\omega_n t}}{\sqrt{T}},
\end{align}
where $\omega_{\rm n}=2\pi n/T$ with an integer $n$. 
This is equivalent to multiplying the integrand by a finite-width Heaviside function, thereby replacing the infinite-time integration with a finite-time window.
The orthogonal condition is given by 
\begin{align}
    \int^{\frac{T}{2}}_{-\frac{T}{2}}dt~\frac{e^{-i\omega_n t}}{\sqrt{T}}\frac{e^{i\omega_n' t}}{\sqrt{T}}=\delta_{n,n'},
\end{align}
where $\delta_{n,n'}$ is the Kronecker delta.
We assume that the two-time correlation function evaluated over a finite measurement time is equal to that in the infinite-time limit,
 $\langle\{\delta\hat{q}_{d\pm}(t), \delta\hat{q}_{d\pm}(t')\}\rangle=\langle\{\delta\hat{q}_{\pm}(t), \delta\hat{q}_{\pm}(t')\}\rangle.$
Based on this assumption, we consider the degree of entanglement \( \langle E_d(\omega_n, T) \rangle \) for a finite measurement time $T$.  
Applying the finite-time Fourier transformation, Eq.~\eqref{E(w)} is reformulated to evaluate the degree of entanglement for a finite measurement time.
\begin{align}
    \langle E_d(\omega_n,T)\rangle=\frac{\langle \hat{R}_{\tilde q_+}(\omega_n,T)^2\rangle\langle \hat{R}_{\tilde p_-}(\omega_n,T)^2\rangle\Omega_-/\Omega}{|\langle[\hat{R}_{\tilde{q}_A}(\omega_n,T),\hat{R}_{\tilde{p}_A}(\omega_n,T)]\rangle|^2}<1.
\end{align}
Since Eq.~\eqref{E(w)} is defined via the two-point correlation function in the Fourier domain, it can be rewritten as the following expression through an appropriate transformation.
\begin{align}
    \langle\{\delta\hat{A}_{d\pm}(\omega_n),\delta\hat{B}_{d\pm}(-\omega_n)\}\rangle&=\int^{\infty}_{-\infty}d\omega\langle\{\delta\hat{A}_{\pm}(\omega),\delta\hat{B}_{\pm}(-\omega)\}\rangle\frac{2\sin{(\frac{T}{2}(\omega_n-\omega))^2}}{T\pi(\omega_n-\omega)^2},\\
    [\delta\hat{A}_{d\pm}(\omega_n),\delta\hat{B}_{d\pm}(-\omega_n)]&=\int^{\infty}_{-\infty}d\omega[\delta\hat{A}_{\pm}(\omega),\delta\hat{B}_{\pm}(-\omega)]\frac{2\sin{(\frac{T}{2}(\omega_n-\omega))^2}}{T\pi(\omega_n-\omega)^2},
\end{align}
where either $A$ or $B$ can be assigned to $q,p$ and $Y$. 
The mathematical relation, $\lim_{T\rightarrow\infty}2\sin^2{(\frac{T}{2}(\omega_n-\omega)})/(T\pi(\omega_n-\omega)^2)=\delta(\omega_n-\omega)$, 
yields $\lim_{T\rightarrow\infty}\langle E_d(\omega_n, T)\rangle=\langle E_\text{Fil}(\omega)\rangle$.
The systematic error of the degree of entanglement can be defined as
\begin{align}
    \Delta E_\mathrm{sys}(\omega_n, T):=\langle E_d(\omega_n, T)\rangle-\langle E_{\text{Fil}}(\omega) \rangle.
\end{align}
The systematic error decreases as the measurement time \( T \) increases.  
It is shown in Appendix~B that the typical timescale at which the systematic error becomes negligible compared to the statistical error is on the order of \( 1/\gamma_m \).  
Accordingly, the systematic error \( \Delta E_{\mathrm{sys}}(\omega_n, T) \) is reduced more rapidly for larger feedback dissipation rates \( \gamma_m \).
\begin{figure}[t]
\begin{center}
\includegraphics[width=90mm]{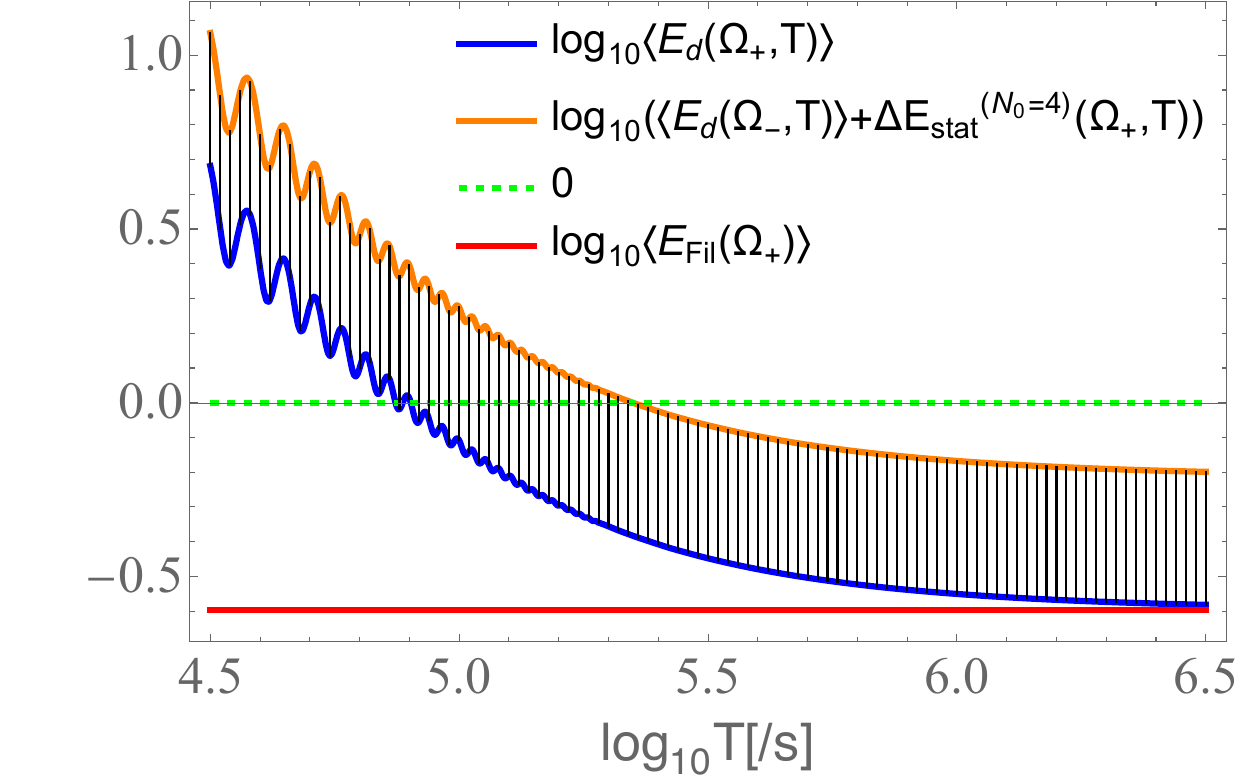}
\hspace{0.5cm}
\includegraphics[width=75mm]{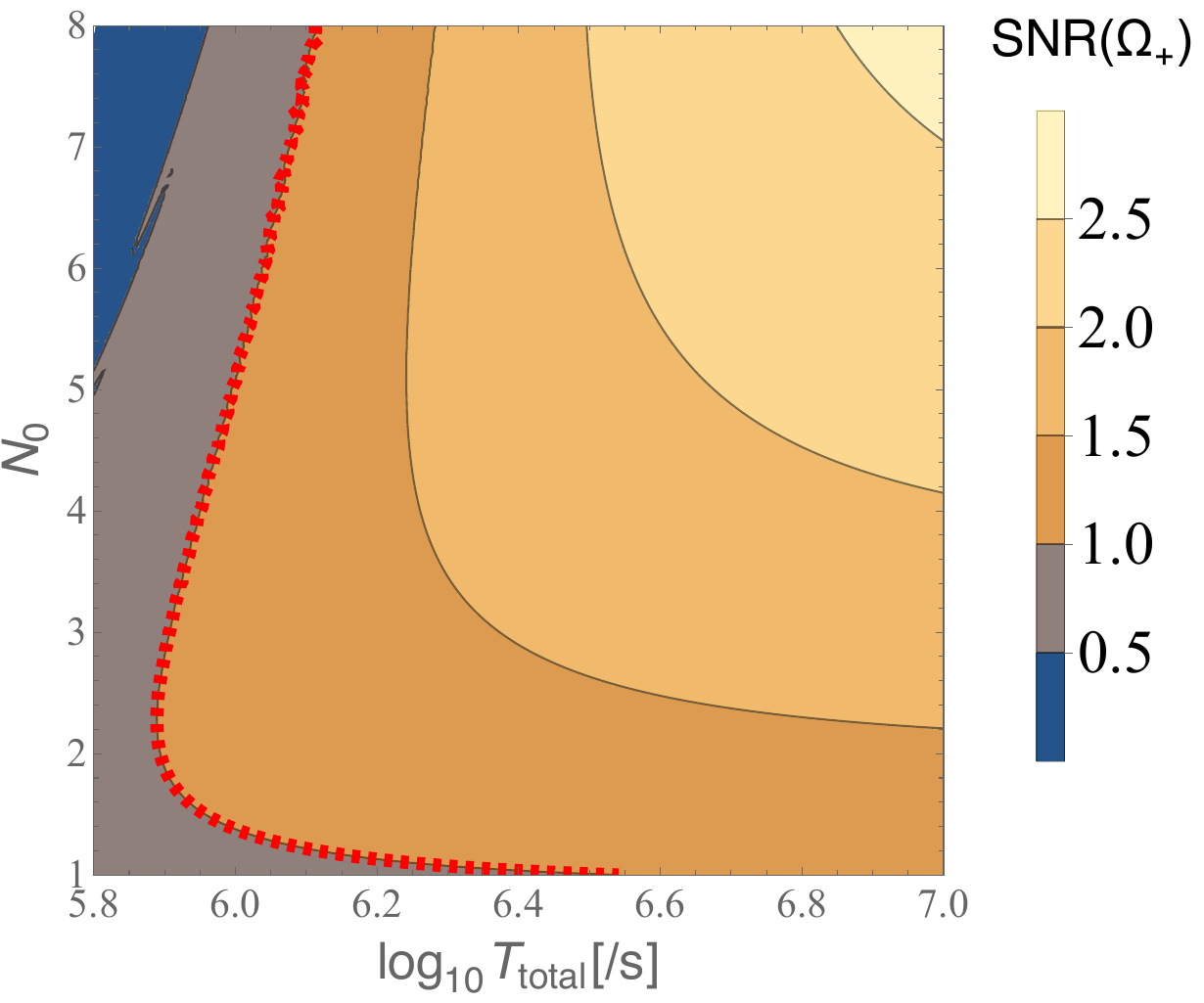}
\caption{
The left panel compares the logarithm of the degree of entanglement and the associated errors as a function of the measurement time \( T \).  
The blue curve represents $\log_{10}$\( \langle E_{d}(\Omega_+, T) \rangle \), which  asymptotically converges to the red line indicating the value of \( \log_{10} \langle E_{\rm Fil}(\Omega_+) \rangle \) in the limit of infinite measurement time.  
The green dotted line shows the boundary of the entanglement criterion, \( \langle E(\Omega_+) \rangle = 1 \), below which the two mirrors are entangled.  
The orange curve represents the degree of entanglement including statistical errors, given by \( \langle E_{d}(\Omega_+, T) \rangle + \Delta E_{\rm stat}^{N_0=4}(\Omega_+, T) \),  
where \( \Delta E_{\rm stat}^{N_0=4}(\Omega_+, T) \) denotes the statistical error for four repeated measurements.  
The black vertical lines
indicate the difference between the blue and orange curves, representing the systematic error for finite measurement time \( T \).
The right panel shows the contour plot of \( \text{SNR}(\Omega_+) \) as a function of the total measurement time \( T_{\text{total}} \) and the number of measurements \( N_0 \).  
The upper-right region above the red dashed curve satisfies \( \text{SNR}(\Omega_+) > 1 \). 
Due to this error, the signal-to-noise ratio (SNR) exceeds unity only when \( T_{\text{total}} > 10^6 \) seconds.  
In this figure, we fix \( r = 1 \), \( \phi = \pi/2 \), $\gamma_m=6.6\times2\pi\times10^{-6}~\mathrm{Hz}$ and 
\( P_{\mathrm{in}} = 10^{-10}~\mathrm{[W]} \).}
\label{lnE-lnT}
\end{center}
\end{figure}

Next, we consider the statistical error of the degree of entanglement.  
The statistical error for \( N_0 \) repeated measurements is given by
\begin{align}
    \Delta E^{(N_0)}_{\rm stat}(\omega_n, T) = \frac{\Delta E^{(1)}_{\rm stat}(\omega_n, T)}{\sqrt{N_0}},
\end{align}
where \( \Delta E^{(1)}_{\rm stat}(\omega_n, T) \) denotes the standard deviation of the degree of entanglement for a single measurement, estimated as
\begin{align}
    \Delta E^{(1)}_{\rm stat}(\omega_n,T)
    &=\sqrt{\langle E_d(\omega_n,T)^2\rangle-\langle E_d(\omega_n,T)\rangle^2}\nonumber\\
    &=\sqrt{\frac{\Big\langle(\hat{R}_{q_{d+}}(\omega_n,T))^4\Big\rangle\Big\langle(\hat{R}_{p_{d-}}(\omega_n,T))^4\Big\rangle\Omega_-^2/\Omega^2}{|\langle\lbrack\hat{R}_{\tilde{q}_{dA}}(\omega)^2, \hat{R}_{\tilde{p}_{dA}}(\omega)^2\rbrack\rangle|^4}-\langle E_d(\omega_n,T)\rangle^2}\nonumber\\
    &=2\sqrt{2}\langle E_d(\omega_n,T)\rangle.
\end{align}
Here, we assume that the measurement outcomes of the mirror operators \( \tilde{q}_\pm(\omega) \) and \( \tilde{p}_\pm(\omega) \) follow Gaussian statistics.  
Specifically, for a Gaussian variable \( Z \) with zero mean value (\( \langle Z \rangle = 0 \)), we use the identity \( \langle Z^4 \rangle = 3 \langle Z^2 \rangle^2 \), as the distribution is fully determined by its variance \( \langle Z^2 \rangle \).  
The statistical error decreases with the number of measurements \( N_0 \), scaling as \( 1/\sqrt{N_0} \) when the measurement time \( T \) is fixed for each individual measurement.

The left panel of Fig.~\ref{lnE-lnT} shows the degree of entanglement and the associated errors as a function of the measurement time $T$.  
The blue curve represents $\log_{10}\langle E_{d}(\Omega_+, T) \rangle$, and the red line indicates the asymptotic value $\log_{10} \langle E_{\rm Fil}(\Omega_+) \rangle$ in the limit of infinite measurement time.  
The green dashed line marks the boundary of the entanglement criterion, $\langle E_\text{Fil}(\Omega_+) \rangle = 1$, below which the two mirrors are entangled.  
The orange curve represents the degree of entanglement including statistical errors, given by 
$\langle E_{d}(\Omega_+, T) \rangle + \Delta E_{\rm stat}^{N_0=4}(\Omega_+, T)$,  
where $\Delta E_{\rm stat}^{N_0=4}(\Omega_+, T)$ denotes the statistical error for four repeated measurements.  
The black vertical lines indicate the difference between the blue and orange curves, representing the systematic error for finite measurement time $T$.

We consider \( N_0 \) repeated measurements, each with a time duration \( T \), such that the total measurement time is \( T_{\text{total}} = N_0 T \).  
The signal-to-noise ratio (SNR) for detecting GIE, based on Eq.~\eqref{E(w)}, is given by 
\begin{align}
    \text{SNR}(\omega_n):=\frac{1-\langle E_d(\omega_n,T_\mathrm{total}/N_0)\rangle}{\Delta E_\mathrm{stat}^{(N_0)}(\omega_n, T_\mathrm{total}/N_0)}.
\end{align}
This provides the simple estimation of the SNR that takes both systematic and statistical errors into account.
The right panel of Fig.~\ref{lnE-lnT} shows the contour plot of \( \text{SNR}(\Omega_+) \) on the plane defined by \( T_{\text{total}} \) and \( N_0 \).
The red dashed line indicates \( \text{SNR}(\Omega_+) = 1 \); the region above this curve satisfies \( \text{SNR}(\Omega_+) \geq 1 \).
For the parameters listed in Table~\ref{table1}, a total measurement time of \( T_{\text{total}} \simeq 10^6~[\mathrm{s}] \) is required to achieve \( \text{SNR}(\Omega_+) = 1 \).
Under these parameters, the statistical error is sufficiently small, and repeating the measurement only $N_0 = 1 \sim 3$ times effectively suppresses statistical fluctuations.
To achieve a higher SNR, it is found that repeating measurements of approximately $10^6$ seconds each is advantageous.
Moreover, reducing the timescale for convergence of the systematic error, \( 1/\gamma_m\), would shorten the time required to reach $\text{SNR}=1$.

In the formulation of the present paper, we assume an idealized weak quantum feedback control without feedback noise.  
When \( \gamma_m \) is significantly increased, feedback noise entering the mirrors can no longer be neglected.
The value of $\gamma_m$ listed in Table I indicates the maximum value for which the feedback noise remains negligible (see Appendix~C). 
A rigorous analytical evaluation of the quantum Wiener filter under quantum feedback control remains a challenging problem and is left for future work.
Furthermore, in the absence of optical squeezing, the value of $\langle E_{\text{Fil}}(\Omega_+)\rangle$ increases, leading to a longer time required to achieve \( \mathrm{SNR} = 1 \).  
For instance, in the case of \( r = 0 \), the required time to reach \( \mathrm{SNR} = 1 \) becomes as long as \( 10^{6.8}~[\mathrm{s}] \), which poses a significant challenge for the practical implementation.
Therefore, the use of squeezed input light plays a crucial role in the GIE detection using the method proposed in the present paper.
\section{Summary and Discussion}
Building upon the results of Ref.~\cite{Miki2024}, we extend the analysis in the Fourier domain to investigate gravity-induced entanglement (GIE) in optomechanical systems using squeezed input light.

First, we considered the effect of a squeezed input light and identified an optimal squeezing angle \( \phi \) for a GIE experiment based on phase measurement.  
At \( \phi \simeq \pi/2 \), we showed that increasing the squeezing strength $r$ effectively reduces optical noise, a mechanism analogous to the use of squeezed input light in gravitational-wave detection, where amplitude noise acting on the mirror is suppressed.
Moreover, by fixing the squeezing angle at $\phi=\pi/2$ in Eq.\eqref{condition3}, we derived the GIE detection criterion, demonstrating that entanglement is generated when the combined effect of gravity-induced coupling and Wiener filtering estimation exceeds the thermal noise, radiation-pressure noise, and measurement noise acting on the mirror.

Second, we estimated the systematic and statistical errors of the degree of entanglement \( \langle E_{\rm Fil}(\omega) \rangle \) under a finite measurement time $T$.  
It was found that the systematic error \( \Delta E_{\rm sys}(\omega_n, T) \) decreases as the measurement time increases, and that its characteristic timescale is proportional to \( 1/\gamma_m \).  
Based on these error estimates, we evaluated the signal-to-noise ratio (SNR) for finite measurement time and determined the total measurement time \( T_{\text{total}} \) required to achieve \( \text{SNR} = 1 \).  
This provides a realistic estimate of the measurement duration necessary to detect and quantify entanglement under finite-time experimental conditions.

In contrast to a previous argument~\cite{Miki2024-2}, which focused on the time required for the system to reach a steady state, our analysis introduces a fundamental timescale that encompasses the entire experimental cycle, including subsequent data analysis, which is necessary for the Fourier analysis.
A realistic approach to GIE detection based on this framework is expected to be further explored in future work.
Future work will involve employing a Wiener filter for accounting for feedback noise, additional noises, and errors in experiments, to enable a more accurate estimation of the time required for GIE detection. 
In particular, models that explicitly include feedback noise may allow for a higher mechanical dissipation rate and thus potentially shorten the detection time. 

Since low-frequency seismic noises  pose a significant challenge for the practical implementation of GIE detection on the earth, space-based experiments are expected to be particularly effective~\cite{matsumoto2025spacebasedmmmgscalelaserinterferometer}. In such environments, in addition to the significant impact of acceleration noise and charging caused by cosmic-ray impacts, the inertial stability of the spacecraft limits the measurement of the common mode. Drag-free control satellites such as LPF have reported thruster noise and fluctuations in external forces \cite{Anderson2018}, and these levels are known to be far larger than those required for GIE detection. 
Therefore, careful optimization of the mirror mass, laser power, and other experimental parameters, together with a passive shielding configuration, is indispensable. 
These remain important directions for future work.
\section{Acknowledgments}
We thank Ryotaro Fukuzumi, Nobuyuki Matsumoto, and Akira Matsumura for valuable discussions and comments on the topic addressed in the present work.  
K.Y. was supported by JSPS KAKENHI Grant No. JP23H01175.  
D.M. was supported by JSPS Overseas Research Fellowships.
\section*{Appendix A}
In this appendix, we show the analytical form for the second-order correlations of $\hat R_{\tilde{q}_+/\tilde p_-}(\omega)$ and the expectation of the commutation relation $\langle[\hat R_{\tilde q_A},\hat R_{\tilde p_A}]\rangle$ to obtain the degree of entanglement $\braket{E_{\rm Fil}(\omega)}$. Using the causal Wiener filter functions in Eqs.~\eqref{Hq} and \eqref{Hp}, we derive
\begin{align}
    \Big\langle\hat{R}_{\tilde{q}_+}(\omega)^2\Big\rangle
    &=\frac{2N_\mathrm{th}+1}{2|F'_+(\omega)|^2}\Bigg(\Omega_+^2\Big(2\gamma_m\frac{2n^+_\mathrm{th}+1}
    {2N_\mathrm{th}+1}+\frac{16g_+^2}{\kappa}(\cosh{2r}+\cos{2\phi}\sinh{2r})\Big)\nonumber\\
    &~~+\frac{\kappa\Big((\omega^{+2}_Y-\Omega^2_+)^2+\omega^2(\gamma^+_Y-\gamma_m)^2\Big)}{16g_\pm^2}(\cosh{2r}-\cos{2\phi}\sinh{2r})+2\Omega_+(\omega^{+2}_Y-\Omega_+^2)\sin{2\phi}\sinh{2r}\Bigg),\label{Rq2}\\
    \Big\langle\hat{R}_{\tilde{p}_-}(\omega)^2\Big\rangle
    &=\frac{2N_\mathrm{th}+1}{2|F'_-(\omega)|^2}\Bigg(\Omega_-^2\Big(2\gamma_m\frac{2n^-_\mathrm{th}+1}{2N_\mathrm{th}+1}+\frac{16g_-^2}{\kappa}(\cosh{2r}+\cos{2\phi}\sinh{2r})\Big)\nonumber\\
    &+\frac{\kappa(\gamma^-_Y-\gamma_m)^2\Omega^4_-+\omega^2(\omega^{-2}_Y-\Omega^2_--(\gamma^-_Y-\gamma_m)\gamma_m)^2}{16g^2_-\Omega^2_-}(\cosh{2r}-\cos{2\phi}\sinh{2r})\nonumber\\
    &+\frac{2\Big(\omega^2(\omega^{-2}_Y-\Omega_-^2-(\gamma^-_Y-\gamma_m)\gamma_m)-(\gamma^-_Y-\gamma_m)^2\Omega^2_-\Big)}{\Omega_-}\sin{2\phi}\sinh{2r}\Bigg).
    \label{Rp2}
\end{align}
We note that the peak value is characterized by $F^\prime(\omega)$ rather than $F(\omega)$ due to the Wiener filtering. In both expressions above, the first line represents the effects of the mechanical thermal noise and radiation pressure noise, while the remaining terms arise from the filtering.
We also obtain the expectation value of the commutator as
\begin{align}
    &|\langle\lbrack\hat{R}_{\tilde{q}_{A}}(\omega),\hat{R}_{\tilde{p}_{A}}(\omega)\rbrack\rangle|^2=\frac{1}{256}\Big|\langle\lbrack\tilde{q}_+(\omega),\tilde{p}_+(-\omega)\rbrack\rangle+\langle\lbrack\tilde{q}_-(\omega),\tilde{p}_-(-\omega)\rbrack\rangle+\langle\lbrack\tilde{q}_+(-\omega),\tilde{p}_+(\omega)\rbrack\rangle+\langle\lbrack\tilde{q}_-(-\omega),\tilde{p}_-(\omega)\rbrack\rangle\Big|^2,\label{RqRp}\\
    &\langle\lbrack\tilde{q}_{\pm}(\omega),\tilde{p}_{\pm}(-\omega)\rbrack\rangle=\frac{1}{|F'_\pm(\omega)|^2}\Bigg(4\gamma_m\omega(\gamma_Y^\pm-\gamma_m+i\omega)-2i\Big(-(\gamma^\pm_Y-\gamma_m)\Omega_\pm^2+i\omega(\omega^{\pm2}_Y-\omega^2_\pm-(\gamma^\pm_Y-\gamma_m)\gamma_m)\Big)\nonumber\\
    &\qquad\qquad\qquad\qquad+2i(\gamma^\pm_Y-\gamma_m+i\omega)(\omega^{\pm2}_Y-\Omega_\pm^2-i\omega(\gamma^\pm_Y-\gamma_m))\Bigg),
    \label{qp}
\end{align}
where we use $\lbrack\hat{p}^\mathrm{in}_{\pm}(\omega),\hat{p}^\mathrm{in}_{\pm}(\omega')\rbrack=({2\omega}/{\Omega_{\pm}})\times2\pi\delta(\omega+\omega')$ and $\lbrack\hat{x}^\mathrm{in}_{\pm}(\omega), \hat{y}^\mathrm{in}_{\pm}(\omega') \rbrack=2i\times2\pi\delta(\omega+\omega')$.
\section*{Appendix B}
In this appendix, we show that the characteristic time required to reduce the systematic error is proportional to \( 1/\gamma_m \).  
The components of \( \langle E_{\rm Fil}(\omega_n, T) \rangle \) are obtained by multiplying the conventional two-point correlation functions—Eqs.~\eqref{Rq2}, \eqref{Rp2}, \eqref{RqRp}, and \eqref{qp}—by a \( T \)-dependent term and integrating over \( \omega \).  
In the residue calculation, the \( T \)-dependent terms can be expressed as a sum of contributions from the integration paths in the upper and lower halves of the complex \( \omega \)-plane,
\begin{align}
    \frac{2\sin{(\frac{T}{2}(\omega_n-\omega))}^2}{T\pi(\omega_n-\omega)^2}=\frac{\exp(-iT(\omega_n-\omega))-1}{-2T(\omega_n-\omega)^2}+\frac{\exp(iT(\omega_n-\omega))-1}{-2T(\omega_n-\omega)^2}.\label{T-dependent}
\end{align}
The integration path for the first (second) term on the right-hand side of Eq.~\eqref{T-dependent} lies in the upper (lower) half of the complex plane.  
Thus, it is found that the exponential decay is governed by the imaginary part of the poles of the integrand.  
The poles of the integrand are located at \( \omega_{\text{pole}} = \omega_n \) and at the zeros of \( F_{\pm}(\omega) \), \( F_{\pm}(\omega)^* \), and \( |F_{\pm}(\omega)|^2 \).The poles from the zeros are explicitly written as 
\begin{align}
    \omega_{\text{pole}}=&
    ~\pm\frac{i\gamma_m\pm\sqrt{-\gamma_m^2+4\Omega_{\pm}^2}}{-2},~{\rm and}~~
    \pm\frac{\Omega_{\pm}}{\sqrt{\Omega_{\pm}^4-\frac{3}{4}\gamma_m^2\Omega_{\pm}^2+\frac{3}{16}\gamma_m^4}}\Bigg(\Omega_{\pm}^2-\frac{\gamma_m^2}{2}\pm i\gamma_m\sqrt{\Omega_{\pm}^2-\frac{\gamma_m^2}{4}}\Bigg)~,
\end{align}
respectively.
Under the assumption \( \Omega_{\pm} \gg \gamma_m \), the imaginary part of \( \omega_{\text{pole}} \) is  \( \gamma_m \).  
Thus, stronger feedback control can shorten the timescale over which the systematic error becomes sufficiently small.

\section*{Appendix C}
Here, the Langevin equation for the perturbed quantities is reformulated to explicitly include a feedback control term~\cite{genes}.
\begin{align}
    \dot{\delta\hat{q}}_{\pm}&=\Omega_{\pm}\delta\hat{p}_{\pm}\\
    \dot{\delta\hat{p}}_{\pm}&=-\Omega_{\pm}\delta\hat{q}_{\pm}-2g_{\pm}\delta\hat{x}_{\pm}-\Gamma\delta\hat{p}_{\pm}+\sqrt{2\Gamma}\hat{P}^\mathrm{in}_{\pm}-\int^t_{-\infty}ds~g_{\pm}(t-s)\hat{Y}_{\pm}(s),
\end{align}
where \( g(t) \) is a causal kernel, which becomes proportional to the derivative of a Dirac delta function in the ideal derivative feedback limit.  
For the phonon noise of the mirrors, we follow the assumption in Ref.~\cite{Miki2024}:
\begin{align}
    \frac{1}{2}\langle\{\hat{P}^\mathrm{in}_{\pm}(t),\hat{P}^\mathrm{in}_{\pm}(t')\}\rangle\simeq\Big(2\frac{k_BT}{\hbar\Omega_{\pm}}+1\Big)\delta(t-t').
\end{align}
Using the Fourier transform and convolution integrals, we obtain the solution to the Langevin equations under feedback control:
\begin{align}
    \delta\hat{q}_{\pm}(\omega)&=\frac{\Omega_{\pm}}{\Omega_{\pm}^2+\alpha_{\pm}\Omega_{\pm}g_{\pm}(\omega)-i\Gamma\omega-\omega^2}\Big(\sqrt{2\Gamma}\hat{P}^\mathrm{in}_{\pm}(\omega)-\alpha_{\pm}\hat{x}^\mathrm{in}_{\pm}+g(\omega)\hat{y}^\mathrm{in}_{\pm}(\omega)\Big),\\
    \delta\hat{p}_{\pm}(\omega)&=-\frac{i\omega}{\Omega_{\pm}}\delta\hat{q}(\omega).
\end{align}
where we define \( \alpha_{\pm} = 4g_{\pm}/\sqrt{\kappa} \), and \( g(\omega) \) is the feedback transfer function.  
A simple and practical choice for \( g(\omega) \) is a standard high-pass filter,
\begin{align}
    g(\omega)=\frac{-i\omega g_\mathrm{cd}}{1-i\omega/\omega_\mathrm{fb}},
\end{align}
where \( g_\mathrm{cd} > 0 \) is the feedback gain, and \( \omega_\mathrm{fb}^{-1} \) represents the time delay of the feedback loop.  
In the ideal limit \( \omega_\mathrm{fb}^{-1} \rightarrow 0 \), the transfer function becomes \( g(\omega) = -i\omega g_\mathrm{cd} \).  
In this case, we have:
\begin{align}
    \delta\hat{q}_{\pm}(\omega)&=\frac{\Omega_{\pm}}{\Omega_{\pm}^2-i(\Gamma+\alpha_{\pm}\Omega_{\pm}g_\mathrm{cd})\omega-\omega^2}\Big(\sqrt{2\Gamma}\hat{P}^\mathrm{in}_{\pm}(\omega)-\alpha_{\pm}\hat{x}^\mathrm{in}_{\pm}-i\omega g_\mathrm{cd}\hat{y}^\mathrm{in}_{\pm}\Big).
    \label{FB-q}
\end{align}
Eq.\eqref{FB-q} implies that the mechanical decay rate under ideal feedback control is given by  
\( \gamma_m = \Gamma + \alpha_{\pm} \Omega_{\pm} g_\mathrm{cd} \), and the feedback noise corresponds to  
\( -i\omega g_\mathrm{cd} \hat{y}^{\mathrm{in}}_{\pm} \).  
When the squeezing parameter \( r > 0 \) and \( \phi = \pi/2 \), the spectral density of \( \delta\hat{q}(\omega) \) is given by:
\begin{align}
    S_{qq}^{\pm}(\omega)=\frac{\Omega_{\pm}^2}{(\Omega_{\pm}^2-\omega^2)^2+(\Gamma+\alpha_{\pm}\Omega_{\pm}g_\mathrm{cd})^2\omega^2}\Big(2\Gamma\Big(2\frac{k_BT+1}{\hbar\Omega_{\pm}}\Big)+\alpha_{\pm}^2(2N_\mathrm{th}+1)e^{-2r}+g_\mathrm{cd}^2\omega_{\pm}^2(2N_\mathrm{th}+1)e^{2r})\Big).
\end{align}
Thus, compared to thermal noise and radiation pressure noise, the feedback noise can be considered negligible under the following condition:
\begin{align}
    2\Gamma\Big(2\frac{k_BT}{\hbar\Omega_{\pm}}+1\Big)+\alpha_{\pm}^2(2N_\mathrm{th}+1)e^{-2r}>g_{cd}^2\omega_{\pm}^2(2N_\mathrm{th}+1)e^{2r},
\end{align}
Using the parameters in Table~\ref{table1}, with \( r = 1 \) and \( \omega = \Omega_+ \), and assuming that the feedback noise is negligible when it is one-tenth of the thermal noise, we obtain upper bounds for \( g_{cd} \) and \( \gamma_m \),
\begin{align}
    g_\mathrm{cd}&\lessapprox\frac{1}{10}\sqrt{\frac{4k_BT\Gamma}{\hbar\Omega_{\pm}^3e^{2r}}}\simeq 0.066,\\
    \gamma_m&=\Gamma+\alpha_{\pm}\Omega_{\pm}g_\mathrm{cd}\simeq6.6\times2\pi\times10^{-6}.
\end{align}
Owing to the practical variation in \( \gamma_m \) between the common and differential modes, \( g_{cd} \) must be individually adjusted for each mode to compensate for the differences.
In this work, we do not incorporate a formalism that explicitly accounts for feedback noise due to the considerable complexity involved in the analytical computation of the quantum causal Wiener filter.

\bibliography{reference}
\end{document}